# Significant enhancement of critical current density in H⁺-intercalated FeSe single crystal

Yan Meng[1,4](孟 炎), Wei Wei[1,4](魏 伟), Xiangzhuo Xing[1,2,3](邢相灼), Xiaolei Yi[1](易晓磊), Nan Zhou[1](周楠), Yufeng Zhang[1](张宇丰), Wenhui Liu[1](刘雯慧), Yue Sun[1,3] (孙悦), and Zhixiang Shi[1,3](施智祥)

[1] *School of Physics, Southeast University, Nanjing 211189, China*

[2] *School of Physics and Physical Engineering, Qufu Normal University, Shandong 273165, China*

[3] Authors to whom any correspondence should be addressed. E-mail: xzxing@qfnu.edu.cn, sunyue@phys.aoyama.ac.jp and zxshi@seu.edu.cn

[4] These authors contributed equally to this paper.

## Abstract

Superconducting transition temperature ($T_c$) and critical current density ($J_c$) are two key factors that are not only crucial for probing high temperature superconducting mechanism, but also for practical applications. The simple crystal structure of FeSe is very favorable for the fabrication of thin films and wires, but its application is limited by the relatively low $T_c$ and small $J_c$. Previous study has found that the $T_c$ of FeSe can be significantly enhanced over 40 K by using protonation method. Here, we present a systematic study of $J_c$ and vortex properties of H⁺-intercalated FeSe (H$_x$-FeSe) single crystals. The value of $J_c$ for H$_x$-FeSe single crystal is significantly enhanced, exceeding $1.3\times10^6$ A/cm² at 4 K, which is more than two orders of magnitude larger than $1.1\times10^4$ A/cm² of pristine FeSe. The vortex pinning mechanism of H$_x$-FeSe is found to be surface pinning, which is different from the dominant strong point-like pinning in pristine FeSe. Moreover, the systematic study of the vortex phase transition and the underlying mechanism provides a wealth of information for the vortex phase diagram of H$_x$-FeSe single crystal. Our results confirm that the introduction of H⁺ intercalations into FeSe not only enhance the $T_c$, but also significantly increase the value of $J_c$, which is favorable for practical applications.

## 1. Introduction

Amongst iron-based superconductors (IBSs), FeSe [1, 2], has attracted considerable interest because of its peculiar properties. One of the significant research subjects in FeSe is about the tunability and improvement of superconducting (SC) transition temperature $T_c$, which is crucial not only for probing high-$T_c$ SC mechanism but also for practical applications. The $T_c$ of the pristine FeSe single crystal is only 9 K [1], while it can be dramatically increased by various methods. Among them, bulk superconductivity of $T_c$ around 30 - 40 K has been realized by intercalating alkaline/alkali-earth metals [3-6], $NH_3$ [7-12], LiOH [13-16], and organic molecules [17-19] into FeSe layers. Another approach to achieve high $T_c$ is based on the ionic liquid gating technique, which can precisely tune the carrier concentrations and improve the $T_c$ of FeSe flakes or films up to 48 K [20, 21]. In addition, unusual discrete SC phases with maximum $T_c$ of 44 K were obtained by a new solid ionic gating device, which can effectively drive $Li^+$/$Na^+$ ions into the bulk FeSe flakes [22].

To date, various methods have been proven to be efficient for achieving high-$T_c$ superconductivity in FeSe. Nevertheless, for the high-$T_c$ FeSe-derived superconductors, $J_c$ and vortex pinning are the most vital properties in practical applications, which have not been widely studied. This is probably due to the limitations of the following unfavorable factors. For instance, both of K-coated and ionic liquid gating methods can only accumulate carriers on the subsurface of FeSe films due to the Thomas-Fermi screening effect [23]. Moreover, traditional gating method is processed by in situ electrical transport measurement at low temperatures with voltage, which brings difficulties to the subsequent post-gating measurements [20]. The instability and sensitivity of metal ions/organic molecules intercalated FeSe also hinder the study of $J_c$ and vortex pinning [24]. Nevertheless, there are still some relevant reports in these high-$T_c$ FeSe-derived superconductors. The largest $J_c$ in ($Li_{1-x}Fe_x$)OHFeSe single crystal reaches a value about $7\times10^5$ A/cm$^2$ at 5 K [25, 26]. Some other studies on $J_c$ have also been carried out in $K_xFe_2Se_2$ [27, 28], Fe(Se,Te) [29] and $Li_x(NH_3)_yFe_2Te_{1.2}Se_{0.8}$ [30] single crystals.

Recently, Cui et al. reported the intercalation of proton ($H^+$) into the bulk materials such as IBSs by a protonation method [31-33], which successfully induces superconductivity or enhances $T_c$. The benefit of such method is that protons are not volatile, hence the protonated samples are relatively stable under ambient conditions, which provides the possibility and convenience for various subsequent measurements. Very recently, we reports a step-like SC phase diagram as a function of the electron concentrations in $H^+$-intercalated FeSe single crystals [34]. The optimal SC phase of $T_c^{onset}$ over 44 K are prepared, which provides an ideal platform to investigate $J_c$ and vortex pinning of FeSe-derived superconductors.

In this study, we systematically studied the enhancement of $J_c$, vortex pinning, and vortex phase diagram in optimally protonated FeSe single crystal with $T_c$ ~ 44 K. The value of $J_c$ for $H_x$-FeSe is significantly enhanced over $1\times10^6$ A/cm$^2$ at 4 K under self-field, which is more than two orders of magnitude higher than $1.1\times10^4$ A/cm$^2$ of pristine FeSe [35, 36]. The pinning mechanism of $H_x$-FeSe can be described by surface pinning, and the vortex glass state of tail region is investigated under the vortex glass theory model. Furthermore, flux pinning behavior and vortex liquid state properties are examined based on the thermally activated flux flow theory (TAFF). The detailed study of the vortex phase transition and the underlying mechanism provides rich information for the vortex phase diagram of $H_x$-FeSe single crystals.

## 2. Experimental details

The pristine FeSe single crystals used for protonation were obtained by the chemical vapor transport growth [35, 37]. The selected high-quality single crystals were then cleaved into thin crystals of approximately 15 μm thickness, with a size of 720 × 410 × 15 μm$^3$. The protonation configuration and process are described in detail in our recently published paper [34]. The optimal $H_x$-FeSe single crystal of $T_\text{c}^\text{onset}$ over 44 K was obtained after 20 days' protonation. The same sample was used for the magnetic and electrical transport measurements. The powder and single-crystal x-ray diffraction results show that H$^+$ intercalation has negligible effect on the lattice structure of $H_x$-FeSe single crystals [34]. Magneto-transport measurements were performed on the Physical Property Measurement System (PPMS-9T) with an excitation current of 1 mA. Magnetic measurements were performed with the VSM option of the Physical Property Measurement System (PPMS-9T). The magnetic field sweeping rate was set to 120 Oe/s for magnetic hysteresis loops.

## 3. Results and discussion

The temperature dependences of resistivity and magnetization measurements are shown in Figs. 1(a) and (b) for pristine FeSe and $H_x$-FeSe single crystal, respectively. It is noted that the normal state resistivity behaviour in $H_x$-FeSe is similar to those of FeSe-derived superconductors such as (Li$_{1-x}$Fe$_x$)OHFeSe [25] and Li$_x$(NH$_3$)$_y$Fe$_2$Se$_2$ [7], but is different from that of pristine FeSe, indicating the change of electronic structure and or scattering mechanism after protonation. The residual resistivity ratio (RRR), defined as $\rho(300\text{ K})/\rho(0\text{ K})$, is about 20 for $H_x$-FeSe, which is smaller than that of pristine FeSe (~70), indicating that more impurity scatterings are introduced simultaneously by H$^+$ intercalations. $H_x$-FeSe exhibits a SC transition temperature $T_\text{c}$ ~ 44 K, which is significantly higher than that of the pristine FeSe of ~9 K. The inset of Fig. 1(b) shows the temperature dependence of field cooling (FC) and zero field cooling (ZFC) magnetization, manifesting that the optimal $H_x$-FeSe single crystal exhibits an almost 100% SC volume, indicating the nature of bulk property.

Magnetic hysteresis loops (MHLs) at different temperatures for FeSe and $H_x$-FeSe superconductors are measured and compared in Figs. 2(a) and (b). The symmetrical MHLs demonstrate the dominance of the bulk pinning in both samples. The second magnetization peaks are absent in all the loops, and $M$ decreases monotonically with the increase of $H$. Such behaviour is quite different from other IBSs [38-42], but is consistent with the FeSe-derived superconductors [35, 36, 43]. The value of $J_\text{c}$ can be derived from the MHLs by the extended Bean model [44], $J_\text{c}=20\Delta M/[a(1-a/3b)]$. Here $\Delta M$ [emu/cm$^3$] is the different value between the magnetization $M_\text{up}$ of upward field sweep and magnetization $M_\text{down}$ of downward field sweep. The $a$ [cm] (width) and $b$ [cm] (length) correspond to the size of the sample ($a < b$). $J_c$ of FeSe and $H_x$-FeSe at different temperatures are illustrated in Figs. 2 (a) and (b), respectively. The derived $J_c$ of $H_x$-FeSe single crystal reaches an extremely large value of $1.33 \times 10^6$ A/cm$^2$ under zero filed at 4 K, which is more than two orders larger than that of the FeSe of $1.1 \times 10^4$ A/cm$^2$ [36, 37]. To the best of our knowledge, such value is one of the largest among FeSe-derived superconductors reported so far, , indicating the great application potential.

In order to comprehensively understand the vortex pinning mechanism, the normalized vortex pinning forces $f = F_p/F_p^{max}$, and the reduced field $h = H/H_{max}$ for FeSe and $H_x$-FeSe under various temperatures are depicted in Figs. 3(a) and (b), correspondingly. Here $H_{max}$ is determined by the peak of the magnetic field, and $F_p^{max}$ represents the maximum pinning force. The $F_p$ was calculated by the formula $F_p = \mu_0 H \times J_c$. Normally in cuprates and IBSs, $h^* = H/H_{irr}$ was commonly used for $f_p$ data calculation [45, 46]. However, considering the difficulty in the determination of $H_{irr}$ in high-$T_c$ superconductors, here $f_p$ data were converted using the reduction field $h = H/H_{max}$. The Dew-Hughes model [47] gives several pinning mechanisms, such as $\Delta\kappa$ pinning, normal point pinning, and surface pinning. There is a certain direct functional relationship between the vortex pinning force density and reduced field $h$ at different temperatures. In a certain temperature range, if $h$ is consistent with one of the below equations, it means that the corresponding pinning mechanism is dominant. Different pinning mechanisms are respectively corresponding to the following equations:

$$f_p = \frac{9}{4}h(1-h/3)^2 \quad \text{for normal point pinning,} \quad (1)$$

$$f_p = \frac{25}{16}h^{1/2}(1-h/5)^2 \quad \text{for surface pinning.} \quad (2)$$

$$f_p = 3h^2(1-2h/3) \quad \text{for } \Delta\kappa \text{ pinning,} \quad (3)$$

As shown in Fig. 3(a), $f_P$ of pristine FeSe can be well fitted by Eq. (1) from 2 to 6 K, suggesting that the pinning mechanism is dominated by the normal point pinning. In contrast, $H_x$-FeSe in Fig. 3(b) is well described by Eq. (2), indicating dominance of the surface pinning, as observed in FeS [48], (Li$_{1-x}$Fe$_x$)OHFeSe [25] and Li$_x$(NH$_3$)$_y$Fe$_2$Te$_{1.2}$Se$_{0.8}$ single crystals [30]. This result suggests that H$^+$ protonation introduces surface-like pinning centers, which results in the enhancement of $J_c$ and $F_p$. It is noteworthy that $f_P$ is affected by flux creep and starts to deviate after $H_{max}$, especially for pristine FeSe. Such deviation implicates that flux creep appears and become dominant at high magnetic fields. Similar results were also found in YBa$_2$Cu$_3$O$_{7-\delta}$ [46] and some other IBSs [45, 49].

Fig. 4(a) presents the variation of resistivity with temperature for $H_x$-FeSe single crystal under different magnetic fields for $H \parallel c$. A rounded feature of transition curve at $T_c^{onset}$ clearly hints the existence of strong SC fluctuation. When the applied magnetic field increases, the resistivity curve at the SC transition continues to extend to lower temperatures, and the SC transition width is getting significantly wider. Moreover, it is clearly seen that the SC transition curve shows a typical foot-like resistivity kink, while the zero resistivity keeps moving toward the low-temperatures as the magnetic field increases. Similar phenomena have also been observed in many other high-$T_c$ superconductors, e.g., cuprates [50-52], (Li,Fe)OHFeSe [14], Li$_x$(NH$_3$)$_y$Fe$_2$Se$_2$ [7], 1111-type [51], and 12442-type IBSs [41]. This foot-like kink has been studied to be directly related to thermal fluctuations, and is responsible for the vortex motion through TAFF. As shown in Fig. 4(b), the characteristic temperature $T_s$ of the foot-like kink is defined as the maximum value of the $(d\ln\rho/dT)^{-1}$ $T$ curve. In high-$T_c$ cuprates, such kink is normally associated with the first-order vortex lattice melting transition [50-63]. The first order transition from Abrikosov vortex lattice to vortex liquid at melting temperature $T_m$ has been theoretically and experimentally demonstrated in a very clean system. When disorders are introduced, the vortex lattice is transformed from vortex liquid to glass. The transition temperature is labelled as $T_g$. Particularly in an intermediately disordered system, Worthington et al proposed that there exists a transitional sluggish vortex liquid region between $T_m$ and $T_g$ [52]. The vortex slush phase has the characteristic of short-range instead of long-range vortex lattice correlation. In H$_x$-FeSe single crystals, the resistive transition curve shows a broaden feature with a typical foot-like resistive tail above zero resistivity, signifying the possible presence of a vortex slush phase with intermediate disorder. It should be noted that both $T_m$ and $T_s$ are the lower temperature limits of the vortex liquid phase boundary. The difference is that $T_m$ is the crossover temperature between the Abrikosov vortex lattice and the vortex liquid in a very clean system, while $T_s$ is the crossover temperature between the vortex slush phase and the vortex liquid in an intermediately disordered system.

In order to understand the state of vortex glass more deeply, we focus on the tail region in the framework of vortex glass theory. On the basis of this theory [55, 56], the resistivity above the glass transition temperature $T_g$ is described by $\rho = \rho_0|T/T_g - 1|^s$, namely the resistivity should decrease with the power law relation. Here, $s$ is a constant associated with the type of disorder, and $\rho_0$ is the characteristic resistivity of the normal state. When the inverse logarithmic derivative of the above equation is taken, $(d\ln\rho/dT)^{-1}$ and $(T - T_g)/s$ should be in a linear relationship. As depicted in Fig. 4(b), the characteristic temperatures $T_g$, $T^*$, and $T_s$ are indicated by arrows. Fitting the linear regions of $(d\ln\rho/dT)^{-1}$ and $T$ gives the values of $T_g$, $T^*$, and the estimated critical exponent $s$, where $T_g$ is the lower temperature limit at which $(d\ln\rho/dT)^{-1}$ is zero. $T^*$ represents the upper temperature limit that deviates from the linear fit. The vortex glass state exists in the region between $T_g$ and $T^*$, and the resistivity curves of H$_x$-FeSe single crystals can be well characterized by the vortex glass state. According to vortex glass theory, the estimated exponent $s$ under different fields should reside in the range of $s \approx$ 2.7~8.5, based on the prediction of the three-dimensional (3D) vortex glass state [55]. For highly anisotropic SmFeAsO$_{0.85}$ single crystals [58], the vortex glass phase exhibits a more pronounced 2D-like behaviour, which possibly originates from the quasi-2D Fermi surface topology.

Next, we investigate the properties of the vortex liquid state above $T_s$ through the magneto-transport measurements, as illustrated in Fig. 4(c). In accordance with the TAFF theory [52, 53, 55, 59], the equation $\rho(T, H) = (2\rho_c U/T)\exp(-U/T) = \rho_{0f} \exp(-U/T)$ gives a good description of the resistivity in the TAFF region, while $U(T, H) = U_0(H)(1-T/T_c)$. By substituting this into the equation above and taking the logarithm of both ends, the formula $\ln\rho(T, H) = \ln\rho_0(H)-U_0(H)/T$ is obtained. Here $U$ is the thermally activation energy, while $U_0(H)$ is the activation energy associated with the magnetic field $H$, which has a hindering effect on the effective pinning. Meanwhile, $\ln\rho_0(H) = \ln\rho_{0f} + U_0(H)/T_c$. Therefore we can simplify the resistivity equation in this region by approximating the Arrhenius relation, $\ln\rho(T, H) = \ln\rho_0(H)-U_0(H)/T$. It can be seen from the above equation that $\ln\rho$ in the TAFF region has a positive linear correlation with $1/T$. The slope of this linear relationship is known to be $-U_0(H)$. According to the variation of resistivity with temperature under different magnetic fields, we depicted the Arrhenius plot of $\ln\rho$ versus $1/T$, as shown in Fig. 4(c). Two significantly different linear TAFF regions are observed in the higher and lower temperature regions, which are labelled by $h$ and $l$, respectively. The intersection of the two linear fitting lines is defined as characteristic temperature $T_{cr}$. Based on the Arrhenius relation, we extract the field dependence of activation energy $U_0(H)$ for the higher temperature region $h$ and the lower temperature region $l$, respectively, which are summarized in Fig. 4(d) in double-logarithmic scales. Obviously, $U_0(H)$ in both temperature regions can be fitted by the power law relation of $U_0(H) \propto H^{-\alpha}$. For region $h$, $U_0(H)$ follows a $\sim H^{-0.54}$ dependence for the whole magnetic fields up to 9 T. However, $U_0(H)$ in the region $l$ is divided into two regions with different values of power exponent $\alpha$ by a crossover field $H \sim 3.6$ T, i.e., $\alpha = 0.66$ for $H < 3.6$ T and $\alpha = 1.3$ for $H > 3.6$ T. It is widely accepted that such a power law dependence with $\alpha = 0.5 \sim 1$ is a characteristic of the 3D vortex matter, whereas a $\ln H$ dependence of the activation energy is often observed in two-dimensional (2D) SC systems [52, 64-66]. Thus, our observations strongly imply the 3D features for the vortices in $H_x$-FeSe single crystal.

The appearance of two TAFF regions in the vortex liquid state has been previously reported in some cuprates superconductors. For example, in YBa$_2$Cu$_3$O$_{7-\delta}$ thin films and YBa$_2$Cu$_4$O$_8$ single crystals [65, 66], Qiu $et$ $al$ observed a crossover in $U_0(H)$ from an $H^{-\alpha}$ behaviour at low temperatures to $\ln H$ at high temperatures, which can be interpreted by the change of the dimensionality of the vortices from a 3D line liquid to a 2D liquid. However, our results reveal that $U_0(H)$ displays an $H^{-\alpha}$ behaviour in both temperature regions $h$ and $l$, so the change of the dimensionality of the vortices is unlikely to account the crossover observed in $H_x$-FeSe single crystal. According to the plastic flow model [67, 68], the vortices are plastically deformed and entangled, which results in the form of $U_0(H) \propto H^{-0.5}$ for the plastic pinning. Hence, the $H^{-\alpha}$ behaviour with $\alpha = 0.54$ in the high temperature region $h$ may be due to the strong plastic pinning. For the low temperature region $l$, $U_0(H)$ displays two regions well separated by a crossover field with $\alpha = 0.66$ for $H < 3.6$ T and $\alpha = 1.3$ for $H > 3.6$ T. Typically, $\alpha \sim 1$ has been commonly observed in some high-$T_c$ cuprates and IBSs [69, 70] and is generally interpreted as an evidence for collective pinning of vortices at high magnetic fields. For $H < 3.6$ T, the value $\alpha = 0.66$ is somewhat higher than that expected for plastic pinning, which may be due to the coexistence of plastic pinning and collective pinning. With increasing magnetic fields, the collective pinning is dominant, corresponding to the value of $\alpha = 1.3$. In addition, it is noted that the magnitude of $U_0(H)$ in the high temperature region $h$ is larger than that of the low temperature region $l$, which has also been observed in some cuprates superconductors [65, 66]. One possible reason is that, as the temperature increases, due to the enhanced thermal fluctuation, the vortex liquid becomes more entangled. Consequently, the entanglement increases the viscosity of vortex liquid and results in a larger $U_0(H)$ in the high temperature region $h$. Certainly, further experiments are needed to verify this interpretation. Moreover, it has also been proposed that the values of power exponent $\alpha$ are intimately related to the type of defects that dominate the flux pinning behaviours. Typically, the planar defects are responsible for $\alpha = 0.5$ and the point defects are responsible for $\alpha = 1$ [41, 55, 64]. From this point of view, the obtained $\alpha = 0.54$ and $\alpha = 0.66$ in the high temperature region $h$ and low temperature region $l$, respectively, are close to 0.5, implying that the planar defects may play an important role on the flux pining, which is consistent with the surface pinning mechanism that is derived from the scaling of pinning force mentioned above. In the low temperature region $l$, we consider that a new kind of pinning center associated with point defects will be formed with increasing magnetic field and result in the larger value of $\alpha = 1.34$ at high magnetic fields.

Finally, based on the results obtained above, we construct the vortex phase diagram for $H_x$-FeSe single crystal, as shown in Fig. 5. The upper critical field $H_{c2}$ is the dividing line between the normal state and the vortex liquid state. The vortex liquid state between $H_{c2}$ and $T_s$ is divided into region I and region II with different dissipation mechanisms governed by TAFF. With decreasing temperature, the vortex liquid-solid phase transition can be classified as several specific regions, consisting of two different kinds of transitions. The first one is called as the vortex slush transition at $T_s$, below which the system with intermediate disorder enters the vortex slush state with short-range lattice correlation. When the temperature is further reduced, a vortex glass critical region between $T^*$ and $T_g$, where the resistivity can be described by the vortex glass theory, appears, and finally the system enters the vortex glass state below $T_g$.

## 4. Summary

To sum up, we systematically studied the $J_c$ and vortex dynamics of $H_x$-FeSe single crystals. $T_c$ of the $H_x$-FeSe single crystal is dramatically increased to 44 K with a bulk SC property. At 4 K, the value of $J_c$ for $H_x$-FeSe is significantly enhanced by two orders of magnitude compared with $1.1 \times 10^4$ A/cm$^2$ of pristine FeSe, exceeding $1.3 \times 10^6$ A/cm$^2$. Comparing the vortex pinning characteristics, the pinning mechanism of $H_x$-FeSe is well described by surface pinning, while pristine FeSe is dominated by strong point-like pinning. The foot-like kink observed in the $\rho$–$T$ curves of $H_x$-FeSe is related to the vortex slush phase, and exhibits short-range correlation of vortex lattice. The flux pinning is governed by varying types of defects at different magnetic fields, which leads to a change in the the activation energies. The detailed study of the vortex phase transition provides abundant information for the vortex phase diagram of $H_x$-FeSe single crystals. High-$T_c$ $H_x$-FeSe single crystals show a significant enhancement of $J_c$, indicating the great application potential.

## Acknowledgements

This work was supported by the National Key R&D Program of China (Grant No. 2018YFA0704300), the Strategic Priority Research Program (B) of the Chinese Academy of Sciences (Grant No. XDB25000000), the National Natural Science Foundation of China (Grants No. U1932217 and No. 11674054).

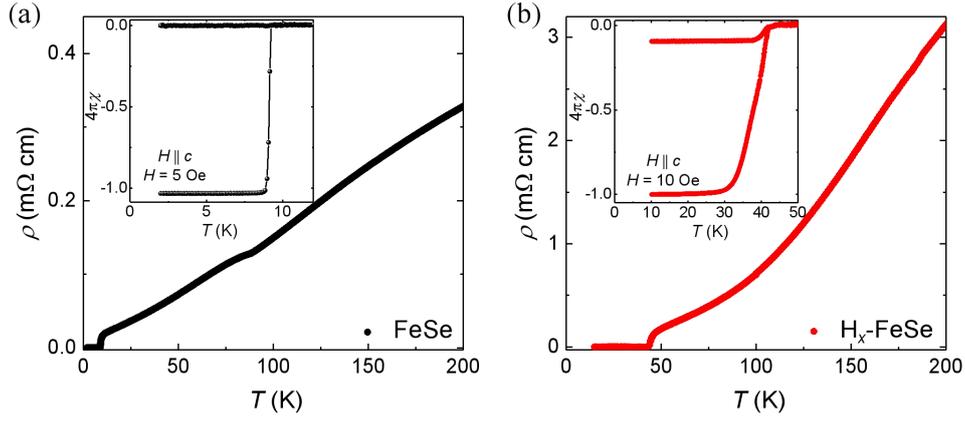

**Fig. 1.** (a) Temperature dependence of the electrical resistivity of FeSe single crystal. The inset shows the temperature dependences of ZFC and FC magnetizations for $H \parallel c$ under 5 Oe. (b) Temperature dependence of the electrical resistivity of $H_x$-FeSe single crystal. The inset shows the temperature dependences of ZFC and FC magnetizations for $H \parallel c$ under 10 Oe.

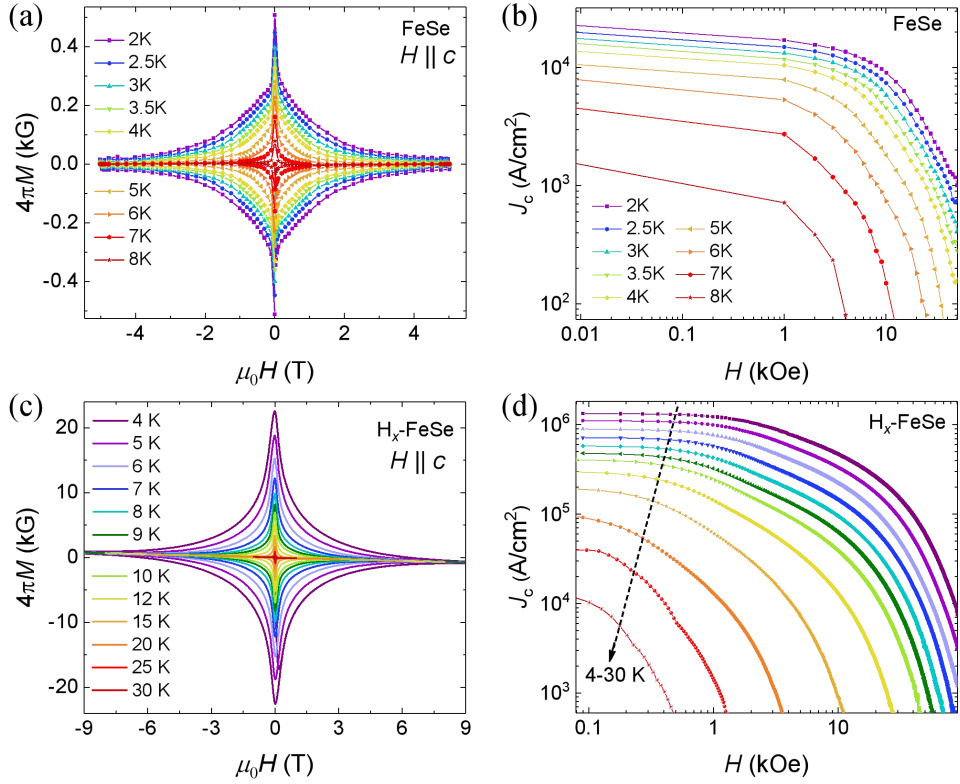

**Figs. 2.** MHLs at different temperatures for $H \parallel c$ and corresponding magnetic field dependence of $J_c$, derived from the Bean model ranging from 2 to 8 K for FeSe single crystal (a and b), and 4−30 K for optimal $H_x$-FeSe single crystal (c and d).

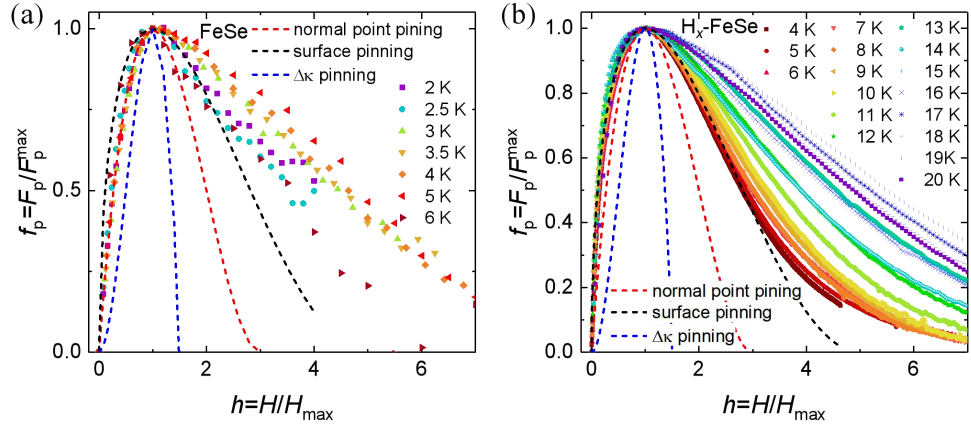

**Fig. 3.** Normalized flux pinning force $f = F_p/F_p^{max}$ as a function of the reduced field $h = H/H_{max}$ at different temperatures for (a) FeSe single crystal (b) H$_x$-FeSe single crystal.

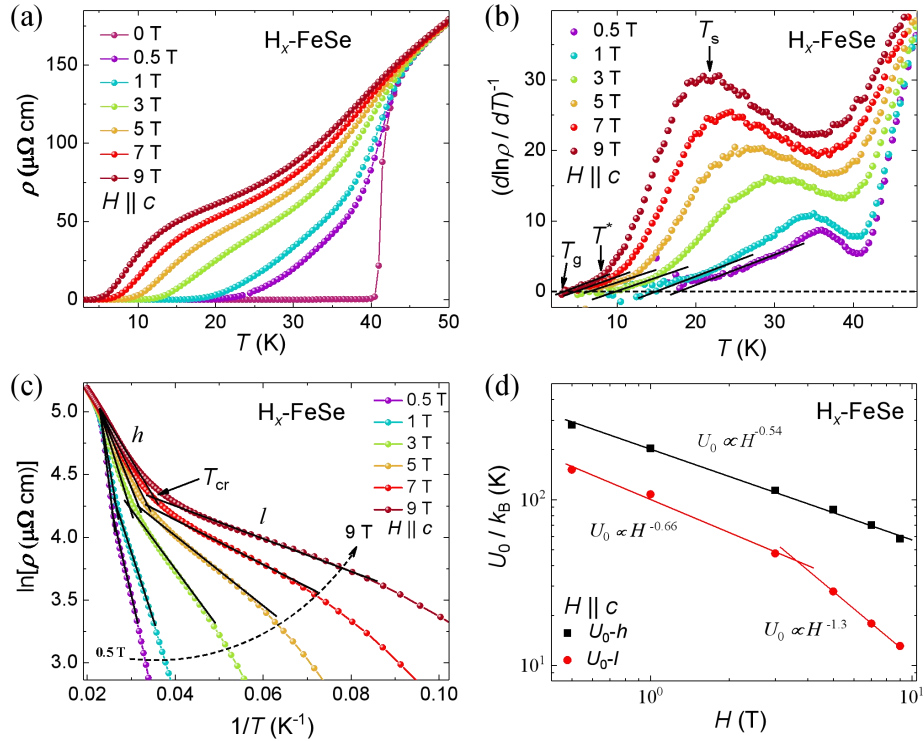

**Fig. 4.** H$_x$-FeSe single crystal (a) Resistivity as a function of temperature for magnetic fields parallel to $c$-axis up to 9 T. (b) Temperature dependence of $(d\ln\rho/dT)^{-1}$ for varying magnetic fields ranging within 9 T for $H \parallel c$, with the associated characteristic temperatures marked with arrows. The dashed line is guide to the eye. (c) Arrhenius plot of the resistive transitions at varying magnetic fields. Two significantly different linear TAFF regions are fitted by black lines in the higher and lower temperature regions, which are labelled by $h$ and $l$ respectively. The intersection of the two linear fitting lines is defined as characteristic temperature $T_{cr}$. (d) The function relation between activation energy $U_0(H)$ and magnetic field $H$ in double-logarithmic scales.

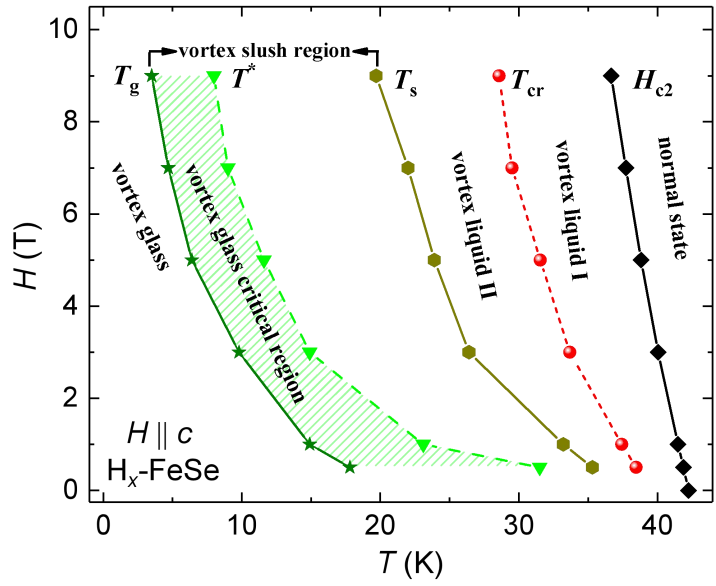

**Fig. 5.** Vortex phase diagram of optimal H$_x$-FeSe single crystal with $H \parallel c$.